\documentclass[10pt,aps,reprint,pra,
    groupedaddress,letterpaper]{revtex4-1}
\usepackage{amssymb}
\usepackage{mathtools}
\usepackage{graphicx}
\usepackage[usenames, dvipsnames, svgnames, table]{xcolor}
\usepackage[colorlinks=true, citecolor=Blue,
            linkcolor=BrickRed, urlcolor=ForestGreen]{hyperref}
\usepackage{txfonts}
\usepackage{mathrsfs}
\usepackage{bm}
\usepackage{booktabs}

\graphicspath{{graphics/}}

\begin{document}

\newcommand{\fixme}[1]{\textcolor{orange}{#1}}
\newcommand{\fixmb}[1]{\textbf{\fixme{#1}}}


\title{Converting the signal-recycling cavity into an unstable optomechanical
filter to enhance the detection bandwidth of gravitational-wave detectors}

\author{Joe Bentley}
\author{Philip Jones}
\author{Denis Martynov}
\author{Andreas Freise}
\author{Haixing Miao}
\affiliation{Institute for Gravitational Wave Astronomy, School of Physics and
Astronomy, University of Birmingham, Birmingham B15 2TT, United Kingdom}
\date{\today}


\begin{abstract}
    Current and future interferometeric gravitational-wave detectors are
    limited predominantly by shot noise at high frequencies.
    Shot noise is reduced by introducing arm cavities and signal
    recycling, however, there exists a trade-off between the peak sensitivity and
    bandwidth. This comes from the accumulated phase of signal sidebands when
    propagating inside the arm cavities. One idea is to cancel such a phase by
    introducing an unstable optomechanical filter. The original design proposed
    in [Phys.~Rev.~Lett.~{\bf 115},~211104 (2015)] requires an additional
    optomechanical filter
    coupled externally to the main interferometer. Here we consider a
    simplified design that converts the signal-recycling cavity itself into the
    unstable filter by using one mirror as a high-frequency mechanical oscillator and introducing
    an additional pump laser. However, the enhancement in bandwidth of this new design is
    less than the original design given the same set of optical parameters.
    The peak sensitivity improvement factor depends on the arm length, the signal-recycling cavity length, and the final detector bandwidth.
	For a 4~km interferometer, if the final detector bandwidth
    is around 2~kHz, with a 20~m signal-recycling cavity, the shot
    noise can be reduced by 10 decibels, in addition to the
    improvement introduced by squeezed light injection.
    We also find that the thermal noise of the mechanical oscillator
    is amplified at low frequencies relative to the vacuum noise,
    while having a flat spectrum at high frequencies.
\end{abstract}


\maketitle


\section{Introduction}

Recent detections of binary black hole (BBH) and binary neutron star (BNS) 
mergers~\cite{Abbott2016a, Abbott2017b, Abbott2017c}
(catalogued in~\cite{Collaboration2018})
have prompted significant research into how the broadband and high-frequency sensitivity of
gravitational wave detectors can be increased. Improving the broadband sensitivity would
increase the signal-to-noise ratio (SNR) of currently visible events, while decreasing the
high-frequency shot noise will allow the determination of the neutron star equation of state~\cite{Read2009,Bauswein2012}.

Current ground-based gravitational-wave (GW) detectors such as
Advanced LIGO~\cite{AdvancedLIGO15shortened} and VIRGO~\cite{AdvancedVirgo15},
as well as future
proposed detectors such as Cosmic Explorer~\cite{Abbott2017shortened} and
Einstein Telescope~\cite{Punturo10shortened} are all limited by the quantum shot
noise at high frequencies.
This is a fundamental noise source that arises due to the intrinsic quantum
uncertainty in the number of
individual photons arriving at the photodetector. Arm cavities are used
to repeatedly reflect the light in the arm cavities, effectively increasing the
path length travelled by the light and amplifying the effect of the
GW strain on its phase. The arm cavities also resonantly
enhance the carrier and increase the intracavity intensity, thereby reducing
the relative photon number uncertainty and thus the shot noise. However, there exists
a strict trade-off between the peak sensitivity and detection bandwidth called
the \emph{Mizuno limit} or \emph{peak sensitivity-bandwidth product}~\cite{phd.Mizuno}
which applies in general to quantum
position measurement
devices using a resonant cavity~\cite{braginsky2000,Tsang2011,Miao2017a}.
This arises due to the positive
dispersion of the arm cavities: when the sideband frequency is near zero the
light is resonantly enhanced by constructive interference, however as the
sideband frequency is increased the light begins to destructively interfere.
The trade-off is also related to the finite storage of energy within the
arm cavities~\cite{Miao2017a}. One approach to improving
the quantum-limited
sensitivity is a direct reduction of the quantum fluctuations using squeezed light
injection~\cite{Caves81,Kimble02,McKenzie04,Oelker16,Schnabel2017,Yap2018,Aggarwal2018},
or internal squeezing~\cite{Rehbein2005,Peano2015,Korobko2017,Mikhail}.
Squeezed light injection will be
included in future upgrades to Advanced LIGO~\cite{Evans2013,Miao13,Miller2015}, as well as in
Cosmic Explorer~\cite{Abbott2017shortened} and Einstein
Telescope~\cite{Hild11shortened}, and has been demonstrated experimentally
in GEO600 and LIGO~\cite{SqueezingNature11,Aasi13}, and as a long-term application
in GEO600~\cite{Grote2013}. However, in light of the Mizuno limit, we will
take another approach, instead broadening the bandwidth of the detector by
introducing a medium with negative dispersion to compensate for the phase
gained in the arm cavity, creating a so-called white light
cavity~\cite{Wicht1997,Wicht2000,Wise2004,Blair2015}.
Previously, atomic systems have been used to
classically demonstrate bandwidth broadening
via negative dispersion~\cite{Pati2007,
Yum2013, Zhou2015, Ma2015}. Another approach is to use
optomechanical coupling such as the unstable
optomechanical filter~\cite{Miao2015} as well as more recent work using
optomechanical resonators~\cite{Zhou2018}. After the bandwidth is broadened, the
Mizuno limit can be used to increase the peak sensitivity by decreasing the
broadened bandwidth.


\begin{figure}[ht]
\includegraphics[width=0.95\linewidth]{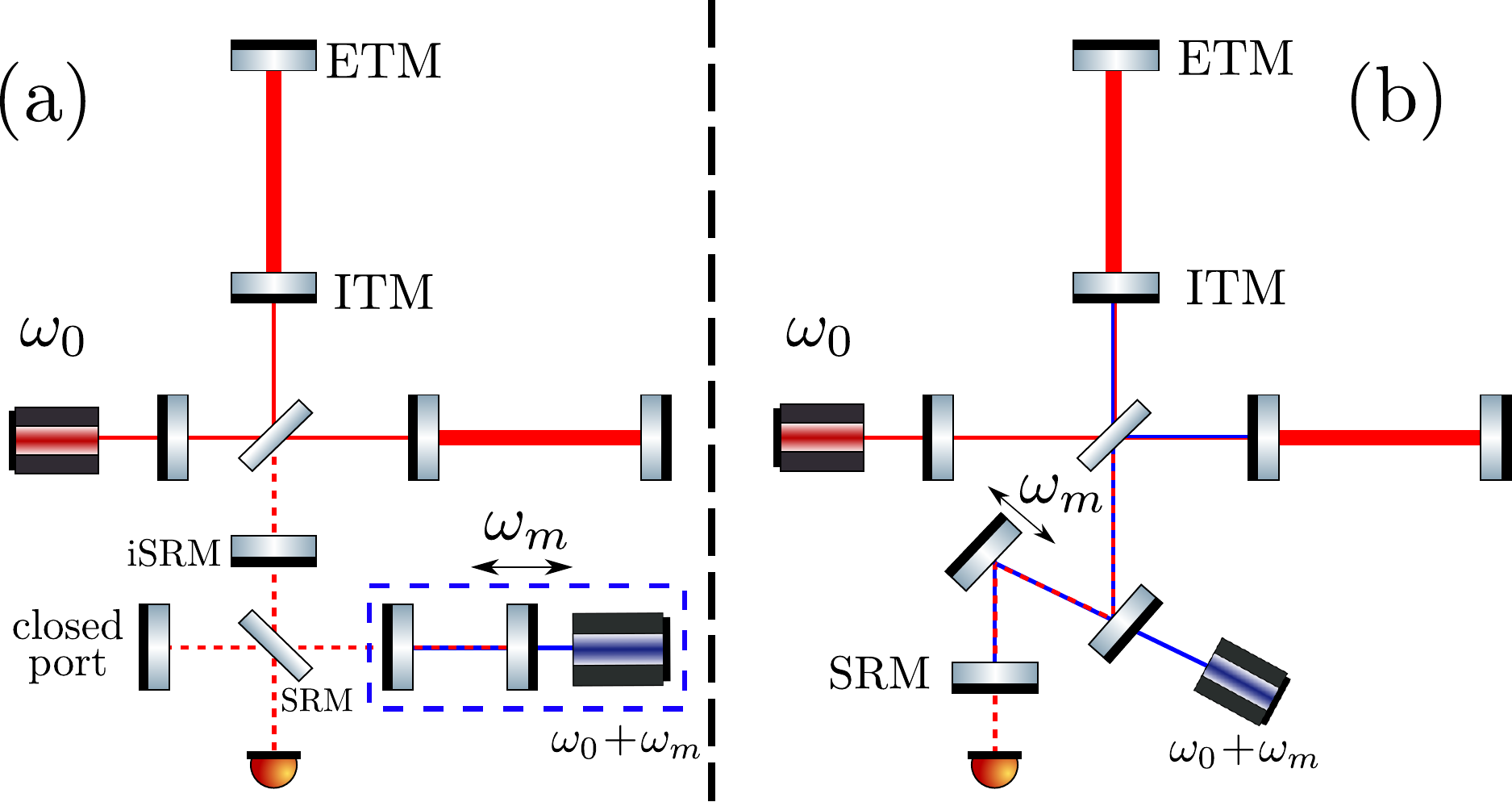}
\caption{\label{fig:setup-comparison}Figure (a) shows a reflection-readout
    design such as in~\cite{Miao2017}.
    The unstable filter alone is highlighted by the blue dashed box.
    Figure (b) shows the new transmission-readout setup proposed in this paper.
    The signal recycling cavity is pumped
    with laser light at \(\omega_0 + \omega_m\), and the mirror above the
    signal recycling mirror is mechanically suspended with mechanical
    resonance frequency \(\omega_m\).
    ETM: End Test Mass, ITM: Input Test Mass, SRM: Signal Recycling Mirror,
    iSRM: Internal SRM (to form an impedance-matched cavity with the ITM)}
\end{figure}

\begin{figure*}
\includegraphics[width=\linewidth]{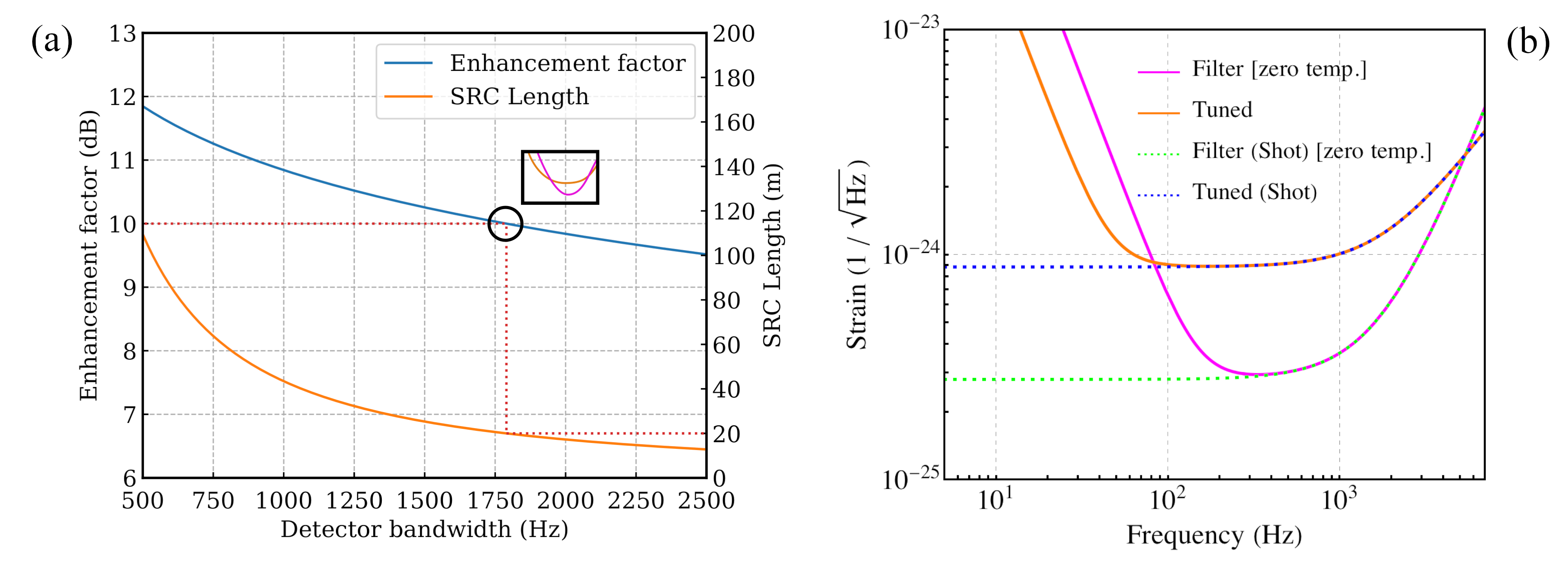}
\caption{\label{fig:results-together}
Figure~(a) shows the peak sensitivity improvement ratio of
the transmission-readout setup to a tuned signal-recycled Michelson interferometer as
a function of broadened detector bandwidth $\Gamma_\text{detector}$ as given in
Eq.\,\eqref{eq:improvement-solved-for-Lsrc}. Here advanced LIGO parameters are used
(4~km arm length, 40~kg test mass, 800~kW arm power), and additionally
an ITM transmissivity of 0.045 and SRM transmissivity of 0.0003 are used.
The SRC length solved for the detector bandwidth is also plotted to
show how it could be varied to improve the peak sensitivity
with these transmissivities.
The dotted line is plotted
for our chosen SRC length of 20~m, giving a detector bandwidth of
around 1790~Hz, and therefore an enhancement factor of
10~dB. The circled value and inset highlights the chosen values used in
figure~(b), which shows the total quantum noise with
the unstable filter mirror at zero temperature.
The tuned Michelson bandwidth set to the effective bandwidth of the new
    transmission readout setup as discussed at the end of
    Sec.~\ref{sec:analysis}. We also
assume 10~dB frequency-dependent squeezing over the entire frequency range as
outlined in~\cite{Kimble02}.}
\end{figure*}

Previous detector designs including an
unstable optomechanical filter (namely~\cite{Miao2015,Miao2017}) have
considered
so-called ``reflection-readout'' based designs. In these designs, the unstable
filter is externally coupled to the main interferometer as shown in
Fig.~\ref{fig:setup-comparison}~(a), requiring drastic alterations of the
detector topology. In this paper, we instead propose a simpler
``transmission-readout'' based design shown
in Fig.~\ref{fig:setup-comparison}~(b), instead
requiring only the conversion of a steering mirror into a high-frequency mechanical oscillator
in the signal
recycling cavity, and the presence of a pump laser at $\omega_0 + \omega_m$.
As an example we apply the design to an example 4km interferometer whose parameters
are described in Fig.~\ref{fig:results-together}, showing that for a detector bandwidth
of 1.8~kHz the shot
noise can be reduced by 10 decibels as shown in
Fig.~\ref{fig:results-together}~(b), however the bandwidth improvement is limited
compared to reflection-readout designs as discussed toward the end of
Sec.~\ref{sec:analysis}. The enhancement factor over a tuned
Michelson as a function of final detector bandwidth is shown in
Fig.~\ref{fig:results-together}~(a).


The outline of this paper is as follows.
In Sec.~\ref{sec:unstable-filter-overview} we will give a brief overview
of the unstable filter originally presented in~\cite{Miao2015},
including its setup and what properties we require of it. In
Sec.~\ref{sec:analysis} we will discuss the new
transmission-readout setup, deriving the quantum noise spectral density
as well as the noise contribution from the thermal noise of the mechanical oscillator.
We will then compare the peak sensitivity improvement achieved over a tuned signal-recycled
Michelson interferometer as shown in Fig.~\ref{fig:results-together},
and discuss the thermal noise of the mechanical oscillator shown in
Fig.~\ref{fig:thermal-noise}. Finally in Sec.~\ref{sec:discussion} we will discuss
how our results compare to previous reflection-readout setups, the thermal
noise and how it can be mitigated with optical dilution, as well as future
developments involving the transmission-readout setup.


\section{\label{sec:unstable-filter-overview}Overview of Unstable Filter}

In this section the motivation and concept of negative dispersion will be
given, followed by a brief overview of the unstable filter as a specific
realization of this concept. For more details on the unstable filter itself
refer to~\cite{Miao2015}.

In a standard tuned Michelson interferometer the arm cavities are tuned to the
carrier light frequency \(\omega_0\) so that \(2\omega_0 L_\text{arm} / c = 2
\pi N \) where \(N\) is an integer, \(L_\text{arm}\) is the arm cavity length,
and \(\omega_0\) is the laser carrier frequency. A GW will induce a change in
path length between the ITM and ETM oppositely for
both arms, modulating the carrier light to induce signal sidebands at
\(\omega_0 \pm \Omega \), with \(\Omega \) being the GW frequency. Since the
arm cavities are tuned to the carrier light these signal sidebands will not be
completely resonant in the arm cavities, as they obtain an extra round-trip
phase
of \(2 \Omega L_\text{arm} / c\) away from the perfect resonant condition which
will
therefore accrue destructively.
As the GW frequency \(\Omega \) increases, as does this extra
round-trip phase, leading to more destructive interference and further reducing
the signal strength. Therefore the arm cavities lead to decreasing sensitivity
at higher frequencies.  We therefore envisage a negative dispersion device that
gives a round-trip phase exactly cancelling that gained in a round-trip through
the arm cavity, so that the round-trip phase gained through this device is \(-2
\Omega L_\text{arm} / c \). This will cancel the attenuation of the signal due
to positive dispersion and effectively broaden the bandwidth of the detector.

The unstable filter, highlighted in Fig.~\ref{fig:setup-comparison},
is just one such
realization of a negative dispersion device. It is an optomechanical device
consisting of a cavity with resonant frequency \(\omega_0\) with a fixed input
mirror and a movable end mirror as a mechanical oscillator with mechanical resonant
frequency \(\omega_m\) and quality factor \(Q_m\); the entire cavity pumped by
laser light at frequency \(\omega_0 + \omega_m\). Signal sidebands at frequency
\(\omega_0 \pm \Omega\) enter the unstable filter and beat with the pump field
at \(\omega_0 + \omega_m\), producing a radiation pressure force fluctuating at
frequency \(\omega_m \pm \Omega\) at the mechanically suspended mirror. This
force moves the mirror which modulates the cavity field to further modify the
sidebands at \(\omega_0 \pm \Omega\), and also modify the mirror's mechanical
motion at frequency \(\omega_m\).  This process is analogous to difference
frequency generation, also known as optical parametric amplification, in
non-linear optics, see for example p.~9 of~\cite{boyd2003nonlinear}. It can be shown that
assuming the so-called resolved sideband regime, where the GW sideband
frequency \(\Omega \ll \gamma_f \ll \omega_m\), and \(\gamma_f\) is the
bandwidth of the filter cavity, and also assuming the system is in the unstable
regime where the mechanical damping rate \(\gamma_m \equiv \omega_m / Q_m\) is
much less than negative damping rate due to the optomechanical interaction
\(\gamma_\text{opt}\), the optical transfer function of the filter cavity takes the form,
\begin{equation}
\frac{\Omega + i \gamma_\text{opt}}{\Omega - i \gamma_\text{opt}}
\approx \exp{\left(-\frac{2i\Omega}{\gamma_\text{opt}}\right)}\,,
\end{equation}
where \(\gamma_\text{opt} \equiv g^2 / \gamma_f\) is the negative
optomechanical damping rate with \(g\)---the optomechanical coupling
strength---as defined later in Eq.\,\eqref{eq:interaction-hamiltonian}. In the
second approximation we assumed that \(\gamma_\text{opt} \gg \Omega\), giving a
\emph{linear} negative dispersion. Clearly, the condition to exactly cancel
the phase gained in the arm cavities is therefore given by,
$\gamma_\text{opt} = c / L_\text{arm}$.

This enhancement is intuitive for the reflection-readout approaches such as in
Fig.~\ref{fig:setup-comparison}~(a), because the signal is recycled with a
negative phase and reinjected into the arm cavity to cancel the positive phase
gained. However this is less intuitive in the transmission-readout approach
shown in
Fig.~\ref{fig:setup-comparison}~(b). Regardless, we will show that
bandwidth broadening is still achieved.

\section{\label{sec:analysis}Analysis}

\begin{figure}
\includegraphics[width=0.8\linewidth]{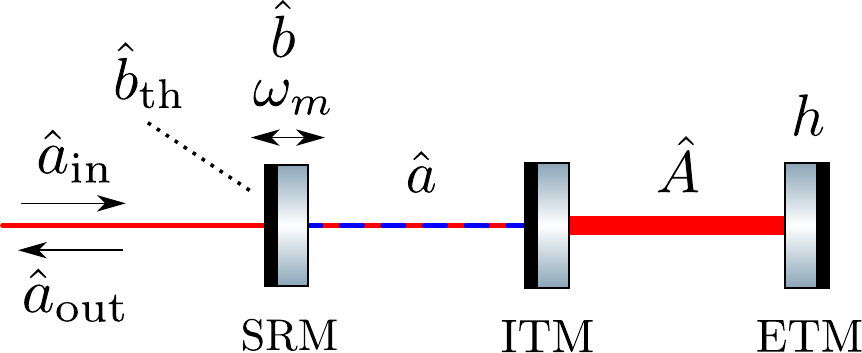}
\caption{\label{fig:setup-analysed}Figure showing the setup analysed in
    Sec.~\ref{sec:analysis}, which is a simplified version of
    Fig.~\ref{fig:setup-comparison}~(b).  \(\hat{a}\) describes the unstable
    filter cavity mode, \(\hat{A}\) describes the differential arm cavity mode,
\(\hat{b}\)
    is the mirror oscillation mode, \(h\) is the GW strain signal, and the
    mirror is coupled to an external heat bath described by the continuum field
    \(\hat{b}_\text{th}\), shown by the dotted line. The cavity field $\hat{a}$
    is coupled to the external continuum fields $\hat{a}_\text{in},
    \hat{a}_\text{out}$.
}
\end{figure}

In this section we outline the analysis of the transmission-readout
setup---a simplified version of which is shown in
Fig.~\ref{fig:setup-analysed} by focusing on the differential mode only.
In comparison to the reflection-readout
setup in Fig.~\ref{fig:setup-comparison}~(a) the arm cavity is no longer
coupled
directly to the dark port. Instead the arm cavity and filter cavity form an
effective three-mirror cavity similar to the twin signal-recycling
scheme studied in~\cite{Thuering2005,Thuering2007}, except with one cavity
replaced with the optomechanical filter cavity.

To analyse the system we use a Hamiltonian-based approach
based on \cite{Chen2013, aspelmeyer14, Law1995}, which was previously
used to analyse the reflection-readout setup in
\cite{Miao2015}. This approach is valid under the \emph{single-mode
approximation}
where the GW sideband frequency \(\Omega / (2\pi) \ll \text{FSR}\) where
\(\text{FSR}\)
is the free spectral range, i.e. only modes within one free spectral range are
considered.
This can be important for long-baseline facilities such as the 40~km
Cosmic Explorer~\cite{Abbott2017shortened} where
the free spectral range is only 3.75~kHz.

The approach consists of first writing the Hamiltonian for the
system, which, referring again to Fig.~\ref{fig:setup-analysed},
consists of SRC mode $\hat{a}$ and differential arm cavity mode
$\hat{A}$, as well as a mechanically suspended
mirror which is modelled as a damped-driven harmonic oscillator with resonant
frequency \(\omega_m\) and mechanical damping rate \(\gamma_m\) which is described
by mode $\hat{b}$. The mirror is additionally coupled to an external continuous mode
$\hat{b}_\text{th}$ which represents an external heat bath.
The equations of motion are then computed using Heisenberg's equation
of motion, and they are then solved in the
frequency-domain to give an equation for an output field at the dark port
\(\hat{a}_\text{out}\) in terms of input fields \(\hat{a}_\text{in}\) and the
GW strain \(h\). The two-photon quadratures~\cite{Caves85,Kimble02}
are then computed, and hence the
spectral density for the quadrature operators.
The external fields $\hat{a}_\text{in}$, $\hat{a}_\text{out}$,
and $\hat{b}_\text{th}$ describe freely propagating continuous mode fields
as in~\cite{BlowLoudonPhoenixEtAl1990}.

The total Hamiltonian is given by \(\hat{H}_\text{tot} = \hat{H}_0 +
\hat{H}_\text{int}^\text{filter} + \hat{H}_\text{int}^\text{ETM} +
\hat{H}_{\gamma_f} + \hat{H}_{\gamma_m} + \hat{H}_{\gamma_\text{arm}}\).
\(\hat{H}_0\) is the free part, \(\hat{H}_\text{int}^\text{filter}\) describes
the interaction between the SRC mode and the mechanically suspended mirror,
\(\hat{H}_\text{int}^\text{ETM}\) describes the radiation pressure coupling
between the ETM and arm cavity field as well as to the GW strain (we choose the
ITM to be static without loss of generality), \(\hat{H}_{\gamma_f}\) describes
the coupling of the SRC to the external continuous vacuum field (the dark port) as
discussed in~\cite{Chen2013} (see footnote \footnote{The method in~\cite{Chen2013} reproduces the quantum Langevin
    equation of~\cite{Gardiner1985} and \cite[p.~158]{Gardiner2010},
    and also~\cite{Clerk2010,aspelmeyer14}, in which the
    setup is known as input-output theory of open quantum systems~\cite{aspelmeyer14}.
    Note that a simple
    cavity coupled to an external bosonic field is mechanically equivalent to a
    mechanical mass coupled to an external heat bath, as evident by comparing
equations~\ref{eq:filter-cavity-eqn-of-motion}
and~\ref{eq:filter-mirror-eqn-of-motion}} for further discussion on this analysis) and
\(\gamma_f = T_\text{SRM} c / (4 L_\text{SRC})\)---where $T_\text{SRM}$ is the
SRM power transmissivity and $L_\text{SRC}$ is the length of the signal recycling cavity---is the coupling constant of the SRC to the external field,
\(\hat{H}_{\gamma_m}\) describes the coupling of the mirror to the external
heat bath \(\hat{b}_\text{th}\) with coupling constant \(\gamma_m
= \omega_m / Q_m\)---where $\omega_m$ is the mirror eigenfrequency and $Q_m$ is the
mechanical quality factor---whose
equation of motion will be the quantum Langevin
equation~\cite{Gardiner1985}~\cite[p.~158]{Gardiner2010}, and finally
\(\hat{H}_{\gamma_\text{arm}}\) describes the coupling between the SRC and
the arm cavity~\cite{Thuering2005,Thuering2007}.

The free Hamiltonian \(\hat{H}_0\) is given by,
\begin{equation}
\hat{H}_0 = \frac{\hat{p}^2}{2 m} + \frac{1}{2} m \omega_m^2 \hat{x}^2 + \hbar
\omega_0 \hat{a}^\dagger \hat{a} + \hbar \omega_0 \hat{A}^\dagger \hat{A} +
\frac{\hat{P}^2}{2 M},
\end{equation}
where \(\hat{P}\) is the ETM momentum, and \(M = m_\text{test} / 4\) is the reduced mass of the
differential
arm cavity mode $\hat{A}$,
and $m_\text{test}$ being the actual mirror mass~\cite{Buonanno02,Buonanno03};
$\omega_0$ is
the carrier frequency of the main interferometer laser and also the resonant frequency
of the arm cavities;
\(\hat{p}\) is
the mechanically suspended mirror momentum, \(\hat{x}\) is its position, and
\(m\) is its mass. 
The filter cavity field $\hat{a}$ is pumped by a laser at frequency
$\omega_0 + \omega_m$ resulting in a
mean photon number $\bar{a}$.
The linearised
filter interaction Hamiltonian is given by~\cite{Chen2013,aspelmeyer14,Miao2015},
\begin{equation}
\hat{H}_\text{int}^\text{filter} = - \hbar g_0 [\hat{a} e^{i (\omega_0 +
\omega_m) t} + \hat{a}^\dagger e^{-i (\omega_0 + \omega_m) t}] \hat{x},
\end{equation}
where \(g_0 = \omega_0 / L_f \bar{a}\), \(\bar{a} = {[2 P_f L_f / (\hbar
\omega_0 c)]}^{1/2}\), where $P_f$ is the circulating power in the filter cavity
and $L_f$ is the length of the filter cavity. The mirror
displacement can be written in the Heisenberg picture
as \(\hat{x} = x_q ( \hat{b} e^{-i \omega_m t} +
\hat{b}^\dagger e^{i \omega_m t})\), where \(x_q\) is the ground-state
harmonic oscillator position uncertainty, \(x_q =
\sqrt{\hbar / (2 m \omega_m)}\). We then move into the rotating frame at
\(\omega_0\) and disregard the \(\omega_0 + 2\omega_m\) sideband by invoking
the rotating wave approximation (RWA), since $\gamma_f \ll \omega_m$ for any
frequency
of interest, however this approximation should be relaxed for a full analysis.
The interaction Hamiltonian therefore becomes,~\cite{Miao2015}
\begin{equation}
\hat{H}_\text{int}^\text{filter} \approx -\hbar g (\hat{a}\hat{b} +
\hat{a}^\dagger \hat{b}^\dagger),
\label{eq:interaction-hamiltonian}
\end{equation}
where \(g = g_0 x_q\).

For the ETM dynamics we have a term due to the effective GW interaction with the mirror, and
another due to the linearised radiation pressure interaction between
the mirror and arm cavity field,
\begin{equation}
\hat{H}_\text{int}^\text{ETM} = F_\text{GW} \hat{X} - \hbar G_0 (\hat{A} +
\hat{A}^\dagger) \hat{X},
\end{equation}
where \(F_\text{GW} = M L_\text{arm} \ddot{h}\) is the GW tidal force, \(G_0 =
\omega_0 / L_\text{arm} \bar{A}\), and \(\bar{A} = {[2 P_\text{arm}
L_\text{arm} / (\hbar \omega_0 c)]}^{1/2}\), where $P_\text{arm}$ is the arm
cavity power, $L_\text{arm}$ is the arm cavity length.

Finally there is sloshing (transfer of excitation) between the SRC and arm
cavity~\cite{Thuering2005,Thuering2007}, which leads to the interaction term,
\begin{equation}
\hat{H}_{\gamma_\text{arm}} = i \hbar \omega_S (\hat{a}\hat{A}^\dagger -
\hat{a}^\dagger \hat{A}),
\end{equation}
where \(\omega_S \approx \sqrt{c \gamma_\text{arm} / L_\text{SRC}}\) is called
the ``sloshing frequency'', with $L_\text{SRC}$ being the length of the
signal-recycling cavity, and
$\gamma_\text{arm} = T_\text{ITM} c / (4 L_\text{arm})$ is the
arm cavity bandwidth, with $T_\text{ITM}$ being the power transmissivity of the ITM.

Therefore we obtain the full set of equations of motion,
\begin{align}
    \hat{a}_\text{out} &= \hat{a}_\text{in} - \sqrt{2 \gamma_f} \hat{a} \\
    \dot{\hat{a}} + \gamma_f \hat{a} &= i g \hat{b}^\dagger + \sqrt{2 \gamma_f}
\hat{a}_\text{in} - \omega_S \hat{A}
    \label{eq:filter-cavity-eqn-of-motion} \\
    \dot{\hat{b}} + \gamma_m \hat{b} &= i g \hat{a}^\dagger + \sqrt{2 \gamma_m}
\hat{b}_\text{th}
    \label{eq:filter-mirror-eqn-of-motion} \\
    \dot{\hat{A}} &= \omega_S \hat{a} + i G_0 \hat{X}
    \label{eq:arm-cavity-eqn-of-motion} \\
    \dot{\hat{X}} &= \hat{P} / M \\
    \dot{\hat{P}} &= - M L \ddot{h} + \hbar G_0 (\hat{A} + \hat{A}^\dagger),
\end{align}
where $\gamma_f$ is the filter cavity bandwidth as defined above.

These equations are then transformed to the frequency domain, noting that
the property of the Fourier transform
$\mathcal{F}\left[\hat{a}^\dagger(t)\right] = (\hat{a}^\dagger)(-\Omega)$ which
we will simply denote as $\hat{a}^\dagger(-\Omega)$.
The equations are then solved to calculate the output field
$\hat{a}_\text{out}(\Omega)$ in terms of the input fields
$\hat{a}_\text{in}(\Omega)$, $\hat{a}^\dagger_\text{in}(-\Omega)$
and the GW strain signal $h(\Omega)$. Note that since $h(t)$ is real,
$h(\Omega) = h^*(-\Omega)$. From these transfer functions
a sideband input-output relation can be constructed of the form,
\begin{equation}
\begin{bmatrix}\hat{a}_\text{out}(\Omega) \\
a^\dagger_\text{out}(-\Omega)\end{bmatrix}
\equiv \mathbb{M}_s
\begin{bmatrix}\hat{a}_\text{in}(\Omega) \\ \hat{a}^\dagger_\text{in}(-\Omega)\end{bmatrix}
+
\mathbb{M}_s^\text{th}
\begin{bmatrix}\hat{b}_\text{th}(\Omega) \\ \hat{b}^\dagger_\text{th}(-\Omega)\end{bmatrix} 
+ \vec{D}_s h(\Omega),
\label{eq:sideband-picture-input-output}
\end{equation}
where $\mathbb{M}_s$ is the transfer matrix of the input field to the output
field at the dark port for the single-photon (sideband) modes, representing the overall
linearised dynamics of the system, similarly $\mathbb{M}_s^\text{th}$ is the transfer matrix
for the thermal noise to the dark port output field, and $\vec{D}_s$ represents
the linearised
coupling of the GW strain signal into the upper and lower sidebands of the dark port
output field.

There is another independent contribution to the noise: the thermal noise
arising from the coupling of the mechanically suspended mirror to the
fluctuating environmental heat bath \(\hat{b}_\text{th}\) at temperature \(T\).
The heat bath provides random thermal fluctuations whose statistics are
determined by the Bose-Einstein distribution, which for $k_B T \gg \hbar
\omega_m$ leads to a spectral density given approximately by~\cite{aspelmeyer14,Miao2015},
\begin{equation}
S_{\hat{b}_\text{th}}(\Omega) = \frac{2 k_B T}{\hbar \omega_m} + 1.
\label{eq:heat-bath-spectral-density}	
\end{equation}

To calculate the power spectral density (PSD) due to the quantum noise
we use the two-photon formalism using quadrature operators $\hat{O}_1,
\hat{O}_2$---respectively called the amplitude and phase
quadratures---with the input quadratures at the dark port having
a flat spectral density equal to unity. These
quadrature operators are related to the single-photon (sideband)
operators by a unitary transformation,
\begin{equation}
\begin{bmatrix}\hat{O}_1\\\hat{O}_2\end{bmatrix}=
\frac{1}{\sqrt{2}}\begin{bmatrix}1&1\\-i&i\end{bmatrix}
\begin{bmatrix}\hat{O}(\Omega)\\\hat{O}^\dagger(-\Omega)\end{bmatrix}
\equiv\mathbb{U}
\begin{bmatrix}\hat{O}(\Omega)\\\hat{O}^\dagger(-\Omega)\end{bmatrix}.
\end{equation}

We need to compute the transfer functions between the output quadratures
and the input quadratures and strain signal of the form,
\begin{equation}
\begin{bmatrix}\hat{a}^\text{out}_1\\\hat{a}^\text{out}_2\end{bmatrix}=
\mathbb{M}_q
\begin{bmatrix}\hat{a}^\text{in}_1\\\hat{a}^\text{in}_2\end{bmatrix}
+\mathbb{M}^\text{th}_q
\begin{bmatrix}\hat{b}^\text{th}_1\\\hat{b}^\text{th}_2\end{bmatrix}
+ \vec{D}_q h(\Omega),
\end{equation}
where $\hat{a}^\text{in}_{1,2}$ and $\hat{a}^\text{out}_{1,2}$ are the
quadratures at the optical input and output port respectively, and
$\hat{b}^\text{th}_{1,2}$ are the quadratures input from the
thermal heat bath. The relation between the quadrature transfer matrices and the sideband
transfer matrices in
Eq.\,\eqref{eq:sideband-picture-input-output} are given by,
\begin{equation}
\mathbb{M}_q = \mathbb{U}\,\mathbb{M}_s\mathbb{U}^\dagger,\quad
\mathbb{M}^\text{th}_q = \mathbb{U}\,\mathbb{M}_s^\text{th}\,\mathbb{U}^\dagger,\quad
\vec{D}_q = \mathbb{U}\,\vec{D}_s.
\end{equation}

The output quadrature operator for a homodyne measurement of homodyne angle
$\zeta$ is given by $\hat{a}^\text{out}_\zeta = (\hat{a}^\text{out}_1,
\hat{a}^\text{out}_2)\cdot(\cos\zeta, \sin\zeta)^T$.
To calculate the spectral density we first separate the output
quadrature into a zero-mean noise term and a mean signal term,
$\hat{a}_\zeta^\text{out}
= \Delta \hat{a}_\zeta^\text{out} + \langle \hat{a}_\zeta^\text{out} \rangle$, where,
\begin{align}
\Delta \hat{a}_\zeta^\text{out} &= \left(
\left[\mathbb{M}_q
\begin{pmatrix}\hat{a}_1^\text{in}\\ \hat{a}_2^\text{in}\end{pmatrix}
\right]^T
+
\left[\mathbb{M}_q^\text{th}
\begin{pmatrix}\hat{b}_1^\text{th}\\ \hat{b}_2^\text{th}\end{pmatrix}
\right]^T
\right)\cdot
\begin{pmatrix}\cos\zeta\\ \sin\zeta\end{pmatrix}
,\\
|\langle\hat{a}_\zeta^\text{out}\rangle|^2
&= |\vec{D}^{(1)}_q \cos\zeta + \vec{D}^{(2)}_q \sin\zeta|^2 |h|^2
\end{align}
where $\vec{D}^{(i)}_q$ is the $i$-th element of $\vec{D}_q$.

The single-sided PSD $S_{OO}(\Omega)$
of an operator $\hat{O}(\Omega)$ for an input
vacuum state $|0\rangle$ is given by the symmetrised 
covariance, $\langle 0|
\hat{O}(\Omega) \hat{O}^\dagger(\Omega')|0\rangle_\text{sym}
= \pi\,S_{OO}(\Omega) \delta(\Omega - \Omega')$~\cite{Kimble02,Braginsky92,Clerk2010}.
First calculating the vacuum noise for $\hat{a}^\text{out}_\zeta$, using that
$\langle 0 | \hat{a}^\text{in}_i(\Omega)\,
	\hat{a}^{\text{in}\,\dagger}_j(\Omega') | 0 \rangle_\text{sym} =
	\pi\,\delta_{ij} \delta(\Omega - \Omega')$,
and then dividing by the strain transfer function, we find the
PSD of the vacuum noise superimposed on the strain measurement is,
\begin{equation}
S_h^\zeta(\Omega) = \frac{(\cos\zeta,\sin\zeta)\mathbb{M}_q(\Omega)\,\mathbb{M}^T_q(\Omega)
(\cos\zeta,\sin\zeta)^T}{|\vec{D}^{(1)}_q \cos\zeta + \vec{D}^{(2)}_q \sin\zeta|^2}.
\end{equation}

We will assume an ideal phase quadrature measurement ($\zeta = \pi / 2$)
at the photodiode,
in which case we are only concerned about the output phase quadrature
$\hat{a}^\text{out}_2$. In this case we have,
\begin{equation}
S_h(\Omega) = \frac{|\mathbb{M}_q^{(2,1)}(\Omega)|^2 + |\mathbb{M}_q^{(2,2)}(\Omega)|^2}{|\vec{D}^{(2)}_q|^2} \equiv S_\text{vacuum}^\text{rp}(\Omega) + S_\text{vacuum}^\text{shot}(\Omega).
\label{eq:vacuum-spectral-density}
\end{equation}

We can follow the same process as above to find the thermal
noise fluctuations arising from thermal noise quadrature operators
$\hat{b}_{1,2}^\text{th}$, noting that the spectral density for the
heat bath is given by Eq.\,\eqref{eq:heat-bath-spectral-density}.

For both the vacuum and thermal noise we define the shot-noise
contributions $S^\text{shot}_\text{vacuum}(\Omega)$
and $S^\text{shot}_\text{thermal}(\Omega)$ as the spectral density
contribution remaining when the mass $M \to \infty$, and
the radiation-pressure contributions $S^\text{rp}_\text{vacuum}(\Omega)$
and $S^\text{rp}_\text{thermal}(\Omega)$ as
the term remaining when the shot-noise contribution is subtracted from the total
spectrum.

\begin{widetext}
Assuming that $\gamma_m \ll \Omega$, we find that the strain-referred
shot-noise limited PSD is given by,
\begin{equation}
	S^\text{shot}_h(\Omega) = \frac{\Omega^2 \gamma _f^2+
\left(g^2-\omega_S^2+\Omega ^2\right)^2}{4 G_0^2 L_\text{arm}^2 \gamma _f
\omega_S^2}
	+\frac{g^2 \gamma _m \left(\frac{2 k_B T}{\hbar  \omega _m}+1\right)}{G_0^2
L_\text{arm}^2 \omega_S^2},
	\label{eq:shot-noise-transmission}
\end{equation}
and the radiation-pressure limited PSD is given by,
\begin{equation}
S^\text{rp}_h(\Omega) =
\frac{4 G_0^2 \hbar^2\omega_S^2}{M^2 \Omega^4 L_{\text{arm}}^2
\left[\Omega^2\gamma_f^2+\left(g^2-\omega_S^2+\Omega^2\right)^2\right]}
\left[\gamma_f + \frac{g^2\gamma_m}{\Omega^2}\left(\frac{2 k_B T}{\hbar\omega_m} + 1\right)\right],
\end{equation}
where in both cases the former term is the vacuum contribution and the latter
term $\propto \gamma_m$ is the thermal contribution.
\end{widetext}

The total strain-referred vacuum-limited PSD can be written in the form,
\begin{equation}
	S_\text{vacuum}(\Omega) \equiv S_h(\Omega)|_{T=0} =
	\left(\frac{1}{\mathcal{K}}+\mathcal{K}\right) \frac{h_\text{SQL}^2}{2}
	\geq h_\text{SQL}^2.
\end{equation}
Here $h_\text{SQL}^2 \equiv 2 \hbar / (M \Omega^2 L_\text{arm}^2) = 8 \hbar / (m_\text{test} \Omega^2 L_\text{arm}^2)$ is the
\emph{standard quantum limit}~\cite{Braginsky92,Kimble02,phd.Miao}, and
$\mathcal{K}$ is a dimensionless factor given by,
\begin{equation}
	\mathcal{K} \equiv \frac{8 P_\text{arm} \omega_0}{M L_\text{arm} c}
	\frac{\gamma_f \omega_S^2}
	{\Omega^2\left(\Omega^2\gamma_f^2+(g^2+\Omega^2-\omega_S^2)^2\right)},
\end{equation}
where the radiation pressure coupling constant $G_0$ has been written out fully.

The transmission-readout shot noise spectral density (given by
Eq.\,\eqref{eq:shot-noise-transmission}) matches peak
sensitivity (shot noise PSD at \(\Omega = 0\)) of a tuned signal-recycled
Michelson interferometer
if we have the condition,
\begin{equation}
g^2 = \omega_S^2 + \gamma_f\omega_S,
\end{equation}
and we set the tuned signal-recycled
Michelson detector bandwidth $\gamma_\text{detector} = \gamma_f$.
In this case the peak sensitivity for both the transmission readout
setup and tuned Michelson is given by,
\begin{equation}
	S^\text{shot}_\text{trans}(\Omega = 0) = S^\text{shot}_\text{tuned}(\Omega = 0) = \frac{\gamma_f}{4 G_0^2 L_\text{arm}^2}.
\end{equation}

The broadened effective detector bandwidth of the transmission readout setup
can be shown to be on the order of
\(\sqrt{\gamma_f \omega_S}\), or in terms of optical parameters,
\begin{equation}
\Gamma_\text{detector} \sim \frac{c}
{2 \sqrt{2}} \left[\frac{T_\text{ITM} T_\text{SRM}^2}{L_\text{arm}
L_\text{SRC}^3}\right]^{1/4}.
\label{eq:gamma-eff}
\end{equation}

To compare the shot noise limited sensitivity of our setup to a tuned
signal-recycled Michelson interferometer,
we set the tuned Michelson detector bandwidth
to be equal to the effective bandwidth of the transmission readout setup, i.e.~
$\gamma_\text{detector} = \Gamma_\text{detector}\,(= \sqrt{\gamma_f \omega_S})$.
In this case the improvement ratio of the
peak power spectral densities, i.e.~the power ratio of the tuned
Michelson to transmission readout
setup shot noise power spectral densities at low frequencies, is given by,
\begin{align}
	\eta &\equiv \frac{S^\text{shot}_\text{tuned}(\Omega = 0)}{S^\text{shot}_\text{trans}(\Omega = 0)}
	= \frac{\sqrt{\gamma_f \omega_S}}{\gamma_f} = \sqrt{\frac{\omega_S}{\gamma_f}} 
\end{align}

\begin{figure}[htb]
\includegraphics[width=\linewidth]{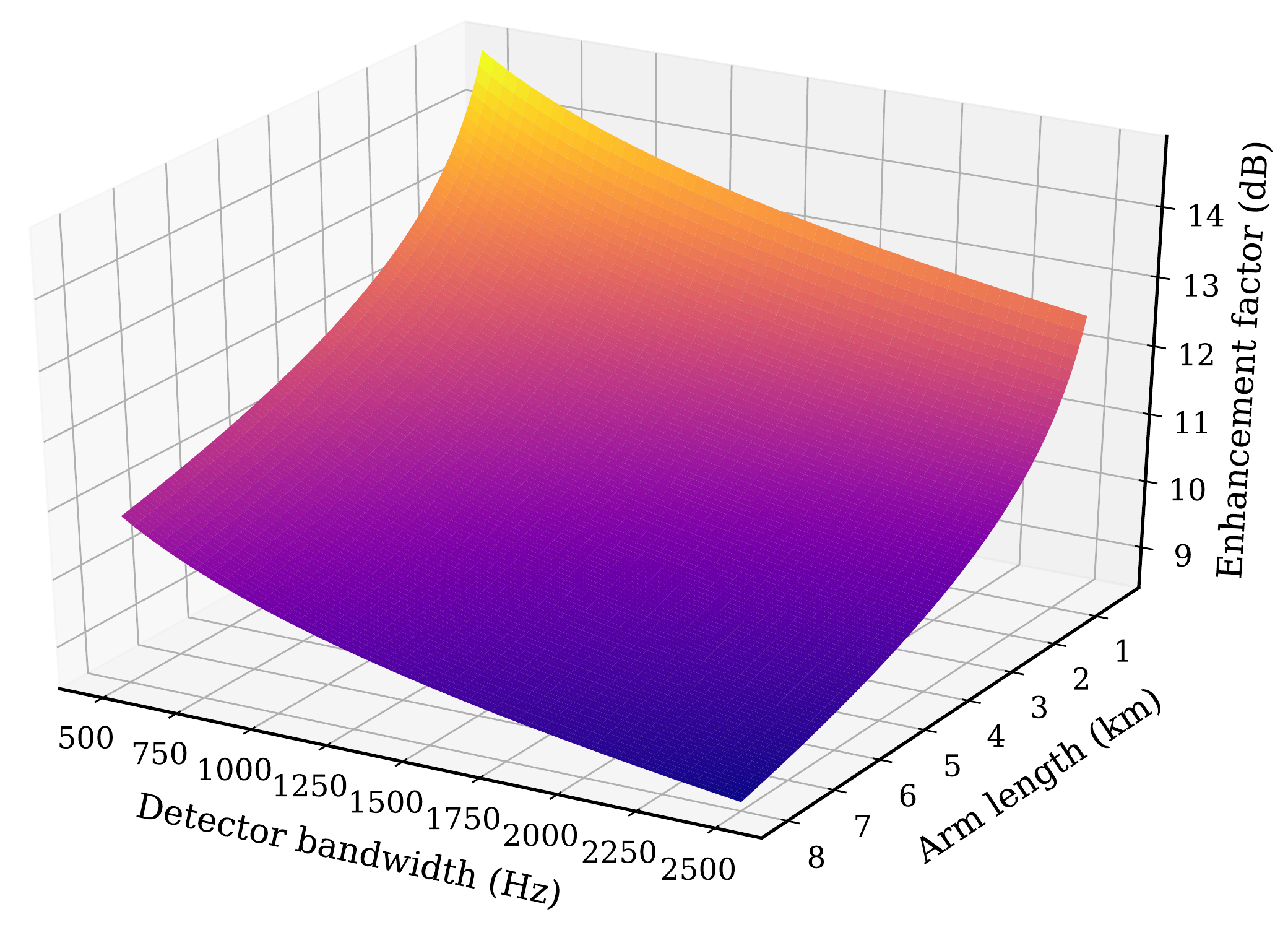}
\caption{
Surface plot showing the peak sensitivity improvement power ratio of
the transmission-readout setup to a tuned Michelson as
a function of both detector bandwidth $\Gamma_\text{detector}$ and
arm length, showing the (log) inverse cube-root dependence of the enhancement
factor on the arm length. The parameters are
as in Fig.\,\ref{fig:results-together}.}
\label{fig:improvement-3d}
\end{figure}

By solving Eq.\,\eqref{eq:gamma-eff} for the SRC length,
the above improvement factor can be written in terms of the effective bandwidth,
arm length, and SRM and ITM power transmissivities as,
\begin{equation}
	\eta = \left(\frac{c T_\text{ITM}}{\Gamma_\text{detector} L_\text{arm} T_\text{SRM}}\right)^{1/3}.
	\label{eq:improvement-solved-for-Lsrc}
\end{equation}
This quantity is shown for various detector bandwidths in
Fig.~\ref{fig:results-together}~(a), and a surface plot for various arm lengths is shown
in Fig.~\ref{fig:improvement-3d}. Note that it is proportional to
$(T_\text{ITM}/T_\text{SRM})^{1/3}$, however if the ITM
transmissivity $T_\text{ITM}$ is increased
then the arm cavity intracavity power will be decreased and hence
the shot noise increased, thereby requiring a higher input power. Similarly
if the SRM transmissivity $T_\text{SRM}$ is very small then losses start to
dominate. Finally, note that the enhancement factor for a given effective detector
bandwidth decreases as the inverse cube root of the arm cavity length $L_\text{arm}$.


\begin{figure}[htb]
\includegraphics[width=0.9\linewidth]{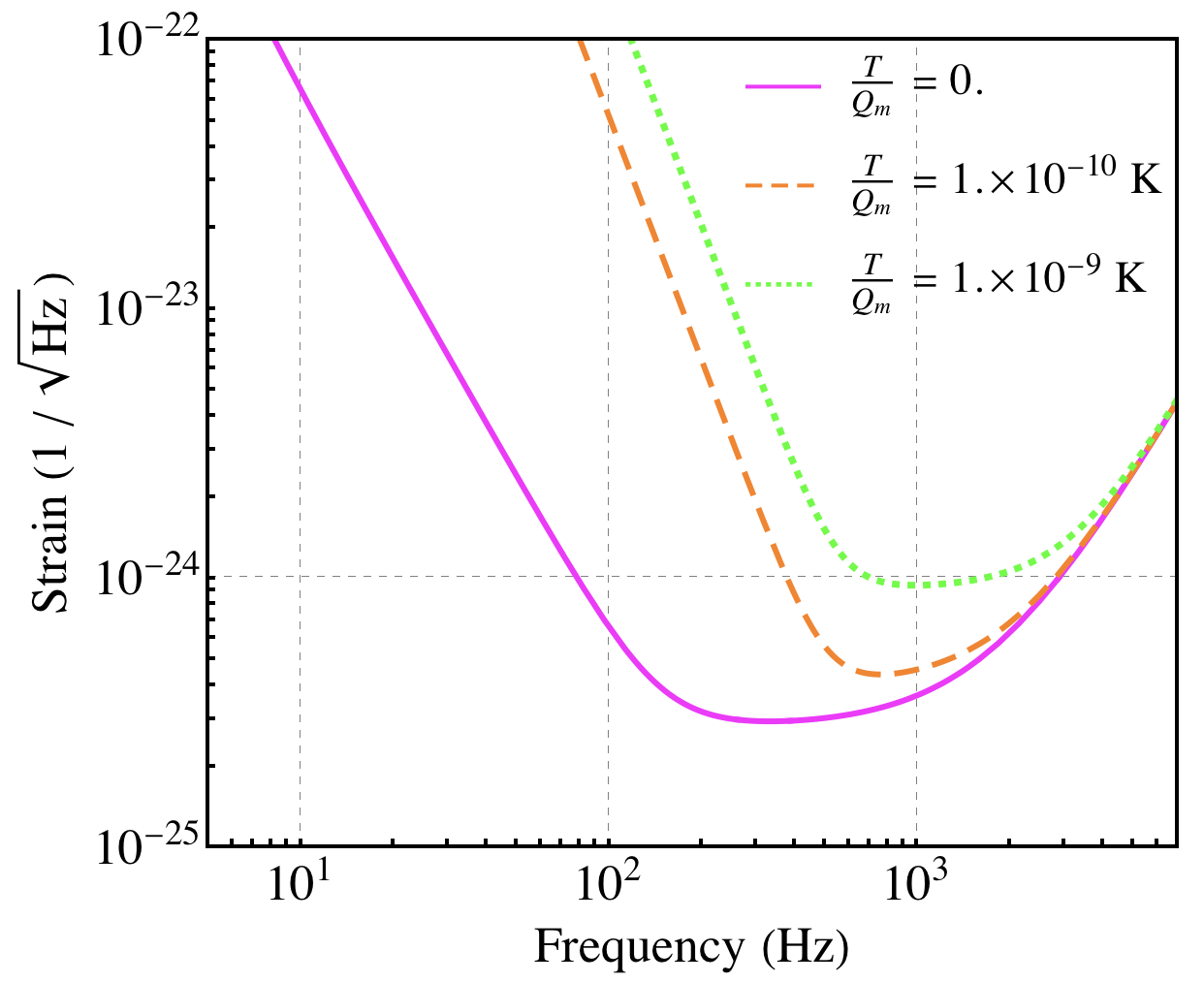}
\caption{\label{fig:thermal-noise}Figure showing the total quantum noise of
	the transmission-readout setup after including the thermal noise at various
	environmental temperatures, using the parameters in Fig.~\ref{fig:results-together},
	including the detector bandwidth marked by the green star.
	At low frequencies the thermal noise is amplified relative
    to the vacuum noise, while at
    high frequencies it has a flat spectrum.}
\end{figure}

For both the shot noise and radiation pressure it was found that the
ratio of the absolute value squared of the thermal fluctuation to the vacuum
fluctuation noise has the form of a ``low-pass filter''. For the shot noise
ratio we assume the resolved sideband regime, whereas for
the radiation pressure ratio no approximation is made. For the shot noise the
effective cutoff frequency is $(g^2 - \omega_S^2) / \gamma_f$. Note
that when the replacement $g^2 = \omega_S^2 + \gamma_f
\omega_S$ is performed, as described later, the cutoff frequency
becomes $\omega_S$. Therefore the thermal noise is suppressed relative
to the vacuum noise at high frequencies where shot noise dominates, however for
$\gamma_m \ll \Omega$ we find that the contribution is approximately flat as shown
in Eq.\,\eqref{eq:shot-noise-transmission}. For the
radiation pressure term, the cutoff frequency $\gamma_m$ is very small compared
to $\Omega$, however the gain $g^2 / (\gamma_f \gamma_m) \equiv \gamma_\text{opt} / \gamma_m$ is very large, and so
at low frequencies the thermal noise is much greater than the vacuum noise.
Intuitively, at low frequencies the thermal heat bath fluctuations are amplified
by the response function of the mechanically suspended mirror in the filter cavity.
The total quantum noise plot is shown in Fig.~\ref{fig:thermal-noise}.
Note that for $\gamma_m \ll \Omega$ the high-frequency thermal noise
contribution is balanced by the diminishing strain response
and has a flat spectrum as shown in Eq.\,\eqref{eq:shot-noise-transmission}.

\begin{figure}[htb]
\includegraphics[width=0.6\linewidth]{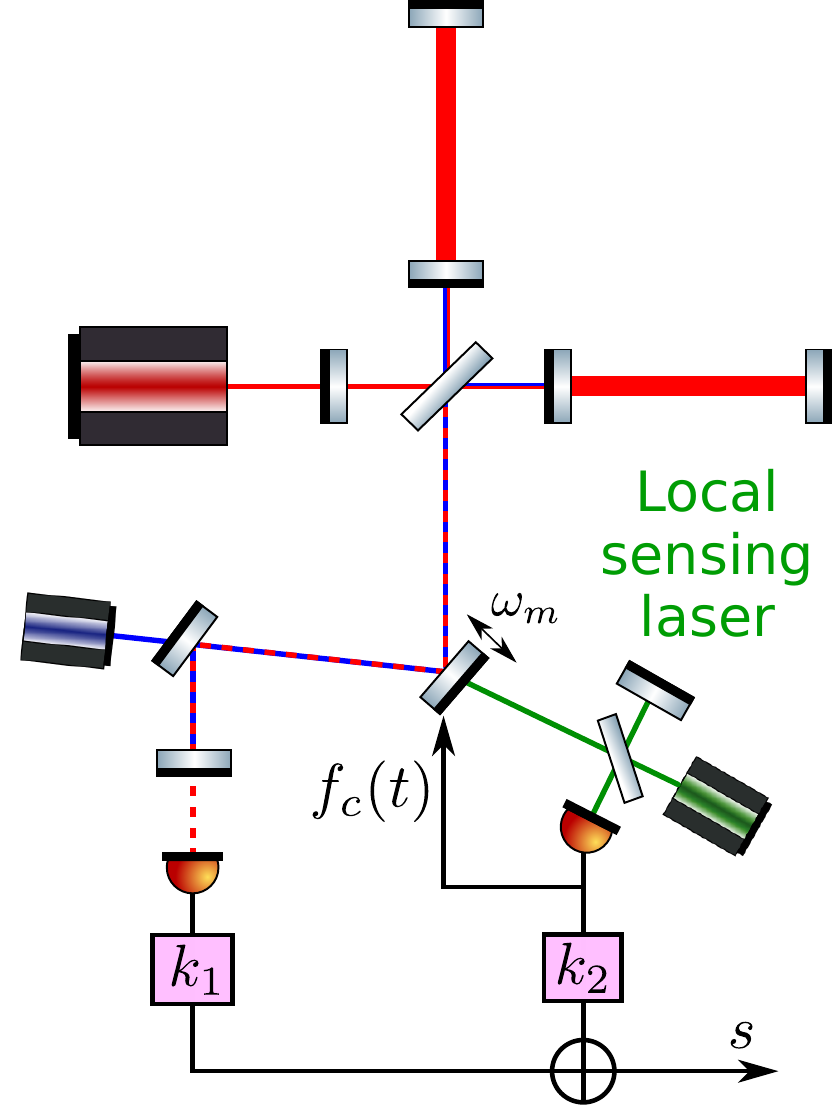}
\caption{\label{fig:local-sensing}Figure showing an example of the local
    sensing control scheme applied to the transmission-readout setup. A control
    force $f_c(t)$ is applied to the mechanically suspended mirror, whose
    displacement is read out by the local sensing laser. The local sensing
readout and main readout are optimally combined by coefficients $k_1$ and $k_2$
to recover the bandwidth broadened sensitivity.}
\end{figure}

\section{\label{sec:discussion}Discussion}



As shown above, the amplitude improvement goes as the square root
of the effective bandwidth, while the decrease in peak sensitivity
of the transmission readout setup goes as the effective bandwidth, and
therefore we are limited in the improvement we can achieve before we start
to degrade the overall sensitivity.
For the previous reflection-readout setups, the effective bandwidth is
given by \((\gamma_f \omega_{S\,\text{refl}}\null^2)^{1/3}\) (where
the sloshing frequency $\omega_{S\,\text{refl}} = \sqrt{c \gamma_f / L_\text{arm}}$ is defined
differently from in our analysis
since we are measuring in reflection of the unstable filter cavity rather than
in transmission~\cite{Miao2015}),
however in that case it can be increased
further by decreasing the filter cavity length or increasing the ETM
transmissivity without adversely affecting the shot noise or radiation pressure
noise.

There is also strong coupling of the thermal noise into the unstable filter,
putting a strict requirement on the environmental temperature. This can be
mitigated using the optical dilution outlined in~\cite{Corbitt07,Chang2012,Ni2012,Korth2013,Reinhardt2016},
stiffening the
dynamics of the suspended mirror, although further R\&D is required
and ongoing to fabricate
mechanical resonators with higher quality
factors via optical dilution or other methods~\cite{Chen2017,Tsaturyan2017,Rossi2018,Page2017,Page2018}.
The thermal noise spectrum in this paper differs from the flat thermal noise
spectrum in previous designs~\cite{Miao2015,Miao2017} since in this case the
thermal heat bath fluctuations are fully shaped by the interferometer, except
coupled indirectly by the mechanically suspended mirror. Overall it was found that at low
frequencies the thermal noise contribution follows the vacuum noise except
it is significantly larger by a factor $\gamma_\text{opt} / \gamma_m \gg 1$, while the
high-frequency thermal noise has a flat contribution.

Another issue is the control of the unstable dynamics of the system.
Previously in~\cite{Miao2015} a stabilizing controller was constructed, however
the time delay of the control signal from the arm cavities to the unstable
filter were neglected. If they are included, it can be shown that the
achievable phase margin will be very small. One other option is to use local
sensing control to locally control the unstable filter, eliminating the time delay. 
Unfortunately, this
will impart significant additional noise on the measurement readout, however as
Denis Martynov has discovered~\cite{DenisLocalSensing} this local sensing noise
can be cancelled out in post-processing by combining the local sensing readout
and main readout optimally. An example of a local sensing control scheme for
the transmission readout setup is given in Fig.~\ref{fig:local-sensing}.

Finally, for the analysis of future GW detectors we will relax the single-mode approximation,
as well as the resolved sideband regime approximation which is manifested in the analysis as
the rotating wave approximation, specifically $\gamma_f \ll \omega_m$. This
will be performed in a followup paper.

\section{Acknowledgements}

We would like to thank members of the LSC AIC, MQM, and QN groups for
fruitful discussions. J.B. is supported by STFC and School of Physics and
Astronomy at the University of Birmingham. J.B., P.J., A.F., D.M., and H.M.
acknowledge the additional support from the Birmingham Institute for
Gravitational Wave Astronomy. H.M. is also supported by UK STFC Ernest
Rutherford Fellowship (Grant No. ST/M005844/11).


\bibliography{bham-ifolab,others}


\end{document}